\documentclass[aps,pre,showpacs,preprintnumbers,amsmath,amssymb,twocolumn,superscriptaddress]{revtex4-1}
\usepackage{graphicx}
\usepackage{dcolumn}
\usepackage{bm}
\usepackage{tabularx}
\usepackage{amssymb}
\usepackage{array}
\usepackage{longtable}
\usepackage{verbatim}
\usepackage{color}
\usepackage{hyperref}

\begin{document}

\title{Extended Vicsek fractals: Laplacian spectra and their applications}

\author{Maxim Dolgushev}
\affiliation {Institute of Physics, University of Freiburg, Hermann-Herder-Strasse 3, D-79104 Freiburg, Germany}
\affiliation {Institut Charles Sadron, Universit\'e de Strasbourg \& CNRS, 23 rue du Loess, 67034 Strasbourg Cedex,
France}

\author{Hongxiao Liu}
\author{Zhongzhi Zhang}
\email{zhangzz@fudan.edu.cn}
\affiliation {School of Computer Science, Fudan University, Shanghai 200433, China}
\affiliation {Shanghai Key Laboratory of Intelligent Information Processing, Fudan University, Shanghai 200433, China}

\begin{abstract}

Extended Vicsek fractals (EVF) are the structures constructed by introducing linear spacers into  traditional Vicsek fractals. Here we study the Laplacian spectra of the EVF. In particularly, the recurrence relations for the Laplacian spectra allow us to obtain an analytic expression for the sum of all inverse nonvanishing Laplacian eigenvalues. This quantity characterizes the large-scale properties, such as the gyration radius of the polymeric structures, or the global mean-first passage time for the random walk processes. Introduction of the linear spacers leads to local heterogeneities, which  reveal themselves, for example, in the dynamics of EVF under external forces.

\end{abstract}
\maketitle

\section{Introduction}\label{intro}

\begin{figure}[t!]
\vspace{0.3cm}
\centering
\includegraphics[width=1\linewidth,trim=0 0 0 0]{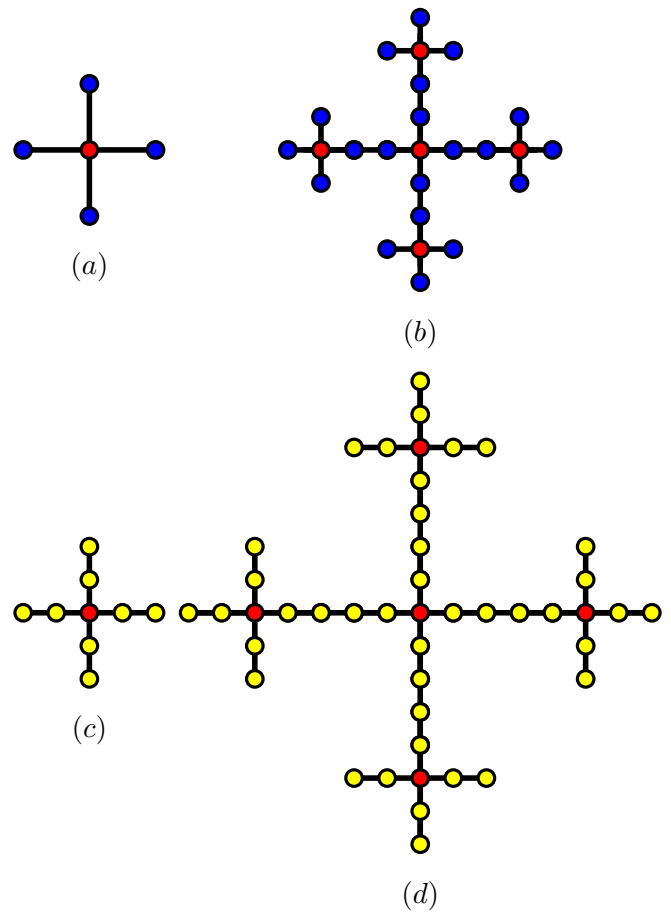}
\caption{ A schematic representation of the construction procedure for extended Vicsek fractals (EVF). An EVF of generation $g$ is constructed from the normal Vicsek fractal (VF)  of generation $g$ by replacing each nonbranching node (see blue beads on (a) for VF of $g=1$ and on (b) for VF of $g=2$) through a linear chain of $k$ nodes (see yellow beads on (c) and (d), here $k=2$). Both for VF and EVF the functionality parameter $f$ denotes the number of nearest neighbors of branching nodes (red beads, here $f=4$).}\label{construction}
\end{figure}

The concept of fractals provides a broadly accepted framework for the modelling of natural phenomena in various fields \cite{mandelbrot83,vicsek92,feder13}. Among different fractal models, the Vicsek fractals (VF) \cite{vicsek83} are a very reliable working horse for the last few decades \cite{jayanthi92,*jayanthi93a,*jayanthi93b,*jayanthi94,schwalm97,blumen03,*blumen04,zhang10,*lin10,*wu12,cherny11,pal12,fan14,fuerstenberg13,fuerstenberg15a,*fuerstenberg15b,kulvelis15,dolgushev15,dolgushev16,vanveen16}.
The popularity of VF is caused due to their full reducibility \cite{jayanthi92,*jayanthi93a,*jayanthi93b,*jayanthi94,blumen03,*blumen04} and also because the VF allow to study different systems obeying distinct scaling laws \cite{blumen03,*blumen04,zhang10,*lin10,*wu12,fuerstenberg13,fuerstenberg15a,*fuerstenberg15b,kulvelis15,dolgushev15,dolgushev16}.
The latter feature is simply achieved based on the same mathematical footing through variation only of the functionality $f$ of branching nodes (i.e., the number of nearest neighbors, see Fig.~\ref{construction}(b) for a VF of $f=4$ and of generation $g=2$). In particular, the VF become very popular in the field of hyperbranched macromolecules, since their Laplacian spectra can be obtained analytically~ \cite{jayanthi92,*jayanthi93a,*jayanthi93b,*jayanthi94,blumen03,*blumen04}: In fact, the Laplacian matrix represents the potential energy of the polymers viewed as beads connected by springs (the so-called generalized Gaussian structures, GGS) and hence it describes for GGS a set of equations of motion \cite{gurtovenko05}. Therefore, the Laplacian spectra carry fundamental information about polymer dynamics \cite{gurtovenko05}.

Nowadays, the advanced synthetic techniques allow to introduce linear spacers into hyperbranched polymers \cite{behera04,jeon07,khalyavina10,segawa13}. Therefore it is of great interest to investigate how such spacers influence the properties of the hyperbranched structures. There are a series of theoretical works that have looked on the spacers' role and the respective analytical treatment of the corresponding Laplacian spectra \cite{eichinger80,kloczkowski89,*erman89,*kloczkowski90,*kloczkowski02,satmarel05,satmarel06}. However, to the best of our knowledge, this question has not been addressed so far to the polymeric fractals. For theoretical scope such extended fractals provide a special interest \cite{jurjiu11,polinska14,sokolov16}, given that they show heterogeneity at different scales.

Let us briefly introduce the structures, on which we focus on here, that we call by "extended Vicsek Fractals" (EVF) in the following. EVF are characterized by three parameters, the generation $g$, the functionality $f$ and the spacers' length $k$. Figure~\ref{construction} illustrates the construction scheme of EVF [parts (c) and (d)] from the conventional VF [parts (a) and (b)]. The conventional VF of generation $g$ is constructed form the VF of generation $g-1$ by replacing each bead of VF of $g-1$ through a star-graph of $f+1$ beads, see the red beads on Fig.~\ref{construction}(b) that have the same structure as all beads of Fig.~\ref{construction}(a). In order to obtain EVF of generation $g$ and spacer parameter $k$, one has to replace all non-branching beads of VF of generation $g$ through linear spacers each of the length $k$. Figure~\ref{construction} exemplifies the procedure for EVF of functionality $f=4$ and spacer parameter $k=2$ for $g=1$ and $2$, where each of the VF's non-branching beads (blue) gets replaced through a chain of $k=2$ beads (yellow). In this way the $(f+1)^g$ beads of a conventional VF yield $N=(kf + 1)(f + 1)^{g - 1}$ beads of EVF.

In this paper we analyze the Laplacian spectra of EVF following the general scheme of Ref.~\cite{satmarel05, satmarel06}. This scheme allow us to analyze the spectra in depth, by looking at static and dynamic properties of the structures in the GGS framework. In particular, we find an analytic expression for a sum of all inverse non-zero Laplacian eigenvalues, which is a fundamental quantity for the gyration radius \cite{forsman76,sommer95,schiessel98,gurtovenko05,jurjiu14}, zero-shear viscosity \cite{ferry80,gurtovenko05}, Wiener index \cite{nitta94}, and global mean-first passage time \cite{montroll69,kozak02,agliari08,zhang10,*lin10,*wu12,benichou14}. The dynamics is considered by looking at the response to external forces.

The paper is structured as follows: In Sec.~\ref{spectra} we apply the methods of Refs.~\cite{satmarel05, satmarel06} to the Laplacian spectra of EVF. In Sec.~\ref{sum} we use these results for calculation of the sum of all inverse nonvanishing eigenvalues of EVF. The properties of the Laplacian spectra are exemplified on the gyration radius at the end of Sec.~\ref{sum} and on the dynamics of EVF under external forces in Sec.~\ref{dynamics}. Finally, Sec.~\ref{conclusions} closes the paper with our conclusions.

\section{Recursion formulae for the Laplacian spectrum.}\label{spectra}

The fundamental object on which we focus here is the Laplacian matrix $\mathbf{L}$. The matrix $\mathbf{L}=(L_{ij})$ characterizes the connectivity of a network by having degrees (in other words, functionalities or number of nearest neighbors) of nodes on the diagonal, $(L_{ii})=f_i$; for two directly connected nodes (say, $i$ and $j$) its elements are $(L_{ij})=(L_{ji})=-1$; all other elements are zero. We note that the Laplacian matrix $\mathbf{L}$ provides a basic means for modelling polymers: In the GGS framework~\cite{gurtovenko05}, where the polymeric structure is represented just by beads connected by harmonic springs, the matrix $\mathbf{L}$ describes the potential energy of the polymer. With this the dynamics of a GGS polymer can be described by a set of linear Langevin equations, that are coupled by $\mathbf{L}$, see Sec.~\ref{dynamics}. In this basic description, however, the effects of hydrodynamics, excluded volume, and bending rigidity are not included.

In order to determine the Laplacian spectra of an EVF, we follow here a general scheme of Ref.~\cite{satmarel05, satmarel06}. This renormalization scheme allows to calculate the Laplacian spectrum of any network that was obtained by replacing each node (having, say, functionality $f$) of the base structure through a symmetric star (having $f$ arms of the same length). The only restriction of the scheme is that the base structure should not posses double bonds. For \textit{treelike} structures (i.e., without loops), the Laplacian spectrum of the resulting structure is split on three classes~\cite{satmarel05, satmarel06}: The first class reflects nontrivial motion of the base structure; hence the calculation of the eigenvalues of the resulting structure requires those of the base one. The second class describes only the motion of terminal linear spacers, while all other beads remain immobile; the resulting eigenvalues describe the motion of a linear chain fixed at one of its ends. In the third class the motion of all branching nodes is characterized by the same direction and amplitude; thus, in the coordinate system associated with the branching nodes (i.e., the center of mass of the base structure) one will observe standing waves in the network. The rigorous procedure leading to these three classes is described elsewhere~\cite{satmarel05}, here we apply it to EVF. Thus,
the Laplacian eigenvalues $\{\lambda ^{(g)}\}$ of an EVF of generation $g$
\begin{enumerate}
\item[(1)] The eigenvalues following from the polynomial equation ${P_{f,k}}({\lambda^{(g)}})={\lambda^{(g-1)}_{\mathrm{base}}}$, where ${P_{f,k}}$ is a ${(2k+1)}$-degree polynomial (\textit{vide infra}) and $\{\lambda^{(g-1)}_{\mathrm{base}}\}$ are all \textit{nonvanishing} eigenvalues of the base VF (i.e., normal VF of the same functionality $f$) of generation $g-1$;
\item[(2)] Roots of the polynomial equation $V_k(1-\lambda/2)=0$, where $V_k$ is the $k$th degree Chebyshev polynomial of the third kind;
\item[(3)] Solutions of the equation ${Q_{f,k}}({\lambda}) = 0$, where ${Q_{f,k}}$ is a $(k+1)$-degree polynomial whose structure is discussed below.
\end{enumerate}

We now turn to consider the three classes listed above. The first class of eigenvalues is generated from its ancestor. As shown in Fig.~\ref{construction}, all branching beads (red) of an EVF of generation $g$ form a normal VF of generation ${g-1}$. Denoting by $\{\lambda_{\textrm{base}}^{(g-1)}\}$ the $(f+1)^{g-1}-1$ nonvanishing eigenvalues of the anterior VF of generation ${g-1}$, the eigenvalues of the first class are obtained from \cite{satmarel05,satmarel06}
\begin{align}\nonumber
{P_{f,k}}(\lambda ) &\equiv f + f{U_{2k - 1}\left(1-\frac{\lambda}{2}\right)} - (f - \lambda ){U_{2k}\left(1-\frac{\lambda}{2}\right)}\\ &= \lambda_{\textrm{base}}^{(g-1)},\label{1stPol}
\end{align}
where $U_i$ is the $i$th degree Chebyshev polynomial of the second kind, for which the following recursive relations hold \cite{mason03}:
\begin{align}\nonumber
{U_0}\left(x\right) &= 1 \\\label{RecU}
{U_1}\left(x\right) &=  2x  \\\nonumber
 {U_i}\left(x\right) &= 2x{U_{i - 1}}\left(x\right) - {U_{i - 2}}\left(x\right) \nonumber
 \end{align}
Thus, ${P_{f,k}}$ is a $(2k+1)$-degree polynomial which leads to $(2k+1)$ eigenvalues for each $\lambda_{\textrm{base}}^{(g-1)}$. Hence the number of eigenvalues in the first class is given by
\begin{equation}\label{N_1}
{N_1} = (2k + 1)[{(f + 1)^{g - 1}} - 1].
\end{equation}

The second class is purely related to the motion of dangling spacers (those linear spacers originating from all $N_{\mathrm{term}}=2+(f-2)(f+1)^{g-1}$ terminal beads of functionality one of the normal VF of generation $g$). Thus, the equation for the eigenvalues of this class do not depend on the fractal nature of EVF and is given by
\begin{equation}\label{eq_2nd_class}
{V_k\left(1-\frac{\lambda}{2}\right)}={U_k\left(1-\frac{\lambda}{2}\right)} - {U_{k - 1}\left(1-\frac{\lambda}{2}\right)} = 0,
\end{equation}
where $V_k(x)$ is the Chebyshev polynomial of the third kind. Based on the roots of $V_k(x)$, $x_i=\cos\frac{(i-1/2)\pi}{k+1/2}$ (see, e.g., Table B.2 of Ref. \cite{mason03}), we obtain then all eigenvalues belonging to the second class:
\begin{equation}\label{lambda2nd}
\lambda_i=4\sin^2\left(\frac{(i-\frac{1}{2})\pi}{2k+1}\right), i=1\dots k.
\end{equation}
The multiplicity $\Delta_g$ of these eigenvalues comes from the number of linearly independent modes related to the motion of dangling chains,
 \begin{equation}\label{egDelta}
\Delta_g = N_{\mathrm{term}}-1=(f-2)(f+1)^{g-1}+1.
 \end{equation}
 Equation \eqref{eq_2nd_class} yields $k$ solutions, each with multiplicity $\Delta_g$. Thus, the total number of the eigenvalues appearing in the second class reads
\begin{equation}\label{N_2}
N_2  = [(f - 2){(f + 1)^{g - 1}} + 1]k.
\end{equation}

The third class represents the situation, when all branching nodes (colored by red on Fig.~\ref{construction}) move in the same direction leading to~\cite{satmarel05,satmarel06}
\begin{align}\nonumber
{Q_{f,k}}(\lambda ) &\equiv \lambda \left[(f - 1){U_{k - 1}\left(1-\frac{\lambda}{2}\right)} + {U_k\left(1-\frac{\lambda}{2}\right)}\right] \\
&=0.
\end{align}
There are $k+1$ solutions arising from the equation above, including the root $0$, i.e.,
\begin{equation}
N_3 = k+1.
\end{equation}

Let us check the total number of eigenvalues that we get from the three classes:
\begin{equation}
{N_1} + {N_2} + {N_3} = (kf + 1){(f + 1)^{g - 1}} = N
 \end{equation}
Thus, the total number eigenmodes (including the translational eigenmode related to the eigenvalue $0$) is exactly equal to the number of beads $N$, i.e., the obtained set of the Laplacian eigenvalues is full.

\begin{table*}
\begin{center}
\begin{tabularx}{1\textwidth}{ c|| XXX | XXX }
\hline\hline
$k$ & $f=3$, Eq.~\eqref{lamda_min_approx} & $f=3$, Eq.~\eqref{1stPol} & difference & $f=4$, Eq.~\eqref{lamda_min_approx} & $f=4$, Eq.~\eqref{1stPol} &  difference \\\hline
$5$ & $1.3928\cdot10^{-11}$ & $1.3983\cdot10^{-11}$ &  $0.40\,\%$ & $1.7539\cdot10^{-12}$ & $1.7579\cdot10^{-12}$ &  $0.23\,\%$ \\
$10$ & $3.7654\cdot10^{-12}$ & $3.7804\cdot10^{-12}$ &  $0.40\,\%$ &$4.7055\cdot10^{-13}$ & $4.7163\cdot10^{-13}$ &  $0.23\,\%$ \\
$20$ & $9.8012\cdot10^{-13}$ & $9.8401\cdot10^{-13}$ &  $0.40\,\%$ & $1.2199\cdot10^{-13}$ & $1.2227\cdot10^{-13}$ & $0.23\,\%$ \\
\hline\hline
\end{tabularx}
\end{center}
\caption{The minimal nonvanishing eigenvalue for EVF of generation $g=10$ and different $f$ and $k$ obtained from the numerical calculations based on Eq.~\eqref{1stPol} and from approximate analytic Eq.~\eqref{lamda_min_approx}.}\label{table}
\end{table*}

We note that the polynomial equation on the eigenvalues of the first class, Eq.~\eqref{1stPol}, enables to find an approximate expression for the minimal nonvanishing eigenvalue (this eigenvalue describes a global antiphase motion of two largest branches connected to the core, see Ref.~\cite{fuerstenberg13}, and therefore it is $(f-1)$-fold degenerate).  Indeed, writing the involved Chebyshev polynomials as
\begin{equation}\label{Cheb_exp}
{U_n}\left(1-\frac{\lambda}{2}\right) = {\alpha _n} + {\beta _n}\lambda + ...
\end{equation}
from the recursion relations for $U_n$'s, Eq.~\eqref{RecU}, we have:
\begin{equation}
\begin{array}{l}
 {\alpha _0} = 1 \\
 {\alpha _1} = 2 \\
 {\alpha _n} = 2{\alpha _{n - 1}} - {\alpha _{n - 2}}
 \end{array}
\end{equation}
and
\begin{equation}
\begin{array}{l}
 {\beta _0} = 0 \\
 {\beta _1} =  - 1 \\
 {\beta _n} = 2{\beta _{n - 1}} - {\alpha _{n - 1}} - {\beta _{n - 2}}
 \end{array}
 \end{equation}
These recursive equations yield
\begin{equation}\label{Cheb_coeff}
\begin{array}{l}
 {\alpha _n} = n + 1 \\
 {\beta _n} =  - \frac{1}{6}({n^3} + 3{n^2} + 2n).
 \end{array}
\end{equation}
Inserting Eqs.~\eqref{Cheb_exp} and \eqref{Cheb_coeff} into Eq.~\eqref{1stPol} and looking at the first order in $\lambda$ we obtain
\begin{equation}
\lambda_{\min} \simeq \frac{\lambda_{\textrm{base,min}}^{(g-1)}}{(2k + 1)(kf + 1)}.
\end{equation}
Here $\lambda_{\textrm{base,min}}$ is the minimal nonvanishing Laplacian of the conventional VF, which can be approximated by
\begin{equation}
\lambda_{\textrm{base,min}}^{(g-1)} \simeq \frac{3(f+1)-\sqrt{9f^2+14f-7}}{2(f + 4)}[3(f+1)]^{3-g},
\end{equation}
see Ref.~\cite{dolgushev16} for details. Thus, the $(f-1)$-fold degenerate minimal nonvanishing Laplacian eigenvalue of EVF follows for $g\geq3$ and $k\geq1$ the approximate expression
\begin{equation}\label{lamda_min_approx}
\lambda_{\min} \simeq  \frac{3(f+1)-\sqrt{9f^2+14f-7}}{2(f + 4)(2k + 1)(kf + 1)}[3(f+1)]^{3-g}.
\end{equation}

Let us now check the performance of the approximate Eq.~\eqref{lamda_min_approx}. In Table~\ref{table} we compare the values of the minimal nonvanishing eigenvalue for EVF of generation $g=10$ and different $f$ and $k$ computed based on Eq.~\eqref{lamda_min_approx} with those coming from the numerical solution of Eq.~\eqref{1stPol}. As can be inferred from the table, Eq.~\eqref{lamda_min_approx} works very well.

\section{The sum of the inverse eigenvalues}\label{sum}

The sum of all inverse, nonvanishing Laplacian eigenvalues $S_g$,
\begin{equation}
S_g\equiv\sum_{i=2}^{N}\frac{1}{\lambda_i^{(g)}},
\end{equation}
where the eigenvalue $\lambda_1^{(g)}=0$ is excluded form the summation, is a key quantity for many fields. The well-known examples are the gyration radius  \cite{forsman76,sommer95,schiessel98,gurtovenko05,jurjiu14}
\begin{equation}\label{Rg}
\langle R_g^2\rangle=\frac{\ell^2}{N}S_g,
\end{equation}
where $\ell^2$ is the mean-square distance between neighboring nodes; the zero-shear viscosity \cite{ferry80,gurtovenko05},
\begin{equation}\label{viscosity}
 \eta_0=\frac{\nu\zeta\ell^2}{6N} S_g,
\end{equation}
where $\nu$ is the monomer density and $\zeta$ is the monomeric friction coefficient;
the Wiener index \cite{nitta94},
\begin{equation}\label{Wiener}
 W=N S_g,
\end{equation}
and the global mean-first passage time \cite{montroll69,kozak02,agliari08,zhang10,*lin10,*wu12,benichou14}
\begin{equation}\label{GMFPT}
 \langle T_g\rangle=2 S_g.
\end{equation}

Summation of the inverse eigenvalues can be performed, based on the Vieta's formulas: Assume that a polynomial
\begin{equation}
\sum\limits_{i = 0}^n {A(i){x^i}}
\end{equation}
has $n$ roots $x_1$, $x_2$, ..., $x_n$, so that
\begin{equation}
\begin{array}{l}
 \prod\limits_{i = 1}^n {{x_i}}  = {( - 1)^n}\frac{{A(0)}}{{A(n)}} \\
 \sum\limits_{i = 1}^n {\frac{{\prod\limits_{j = 1}^n {{x_j}} }}{{{x_i}}}}  = {( - 1)^{n - 1}}\frac{{A(1)}}{{A(n)}}.
 \end{array}
\end{equation}
Combining the above expressions leads to
\begin{equation}\label{Vieta}
 \sum\limits_{i = 1}^n {\frac{1}{x_i}}  = -\frac{{A(1)}}{{A(0)}}.
\end{equation}

Since the eigenvalues are divided on three classes, the sum of their inverse will be performed in three steps. First, we examine the first class, by considering the polynomial
\begin{equation}\label{1stPol2}
{P_{f,k}}(\lambda) - \lambda _{\textrm{base}}^{(g-1)},
\end{equation}
see Eq.~\eqref{1stPol}. Using the expansion of Eq.~\eqref{Cheb_exp} and denoting the $k$th degree coefficients of the $\lambda^k$ in the polynomial of Eq.~\eqref{1stPol2} by $A^{(1)}(k)$, we have:
\begin{equation}
\begin{array}{l}
A^{(1)}(0) =  - \lambda _{\textrm{base}}^{(g-1)} \\
A^{(1)}(1) = (2k + 1)(kf + 1),
\end{array}
\end{equation}
from which, based on Eq.~\eqref{Vieta}, the sum of the inverse roots for the first class is given by
\begin{equation}
\sum\limits_{i = 1}^{2k + 1} \frac{1}{{{\lambda _i}}} = \frac{{(2k + 1)(kf + 1)}}{\lambda_{\textrm{base}}^{(g-1)}}
\end{equation}

The sum of the inverse eigenvalues in normal VF has been obtained in Ref.~\cite{zhang10},
\begin{align}\nonumber
{\Lambda _g} &\equiv\sum_{i=2}^{(f+1)^g}\frac{1}{(\lambda_{\textrm{base}}^{(g)})_i}\\ &= \frac{{(f - 2){{(f + 1)}^{g - 1}}({3^g} - 1)}}{2} + \frac{{f + 2}}{{f + 1}}\frac{{{3^g}{{(f + 1)}^g} - 1}}{{3f + 2}}.
\end{align}
Then, for EVF, the sum of the inverse eigenvalues belonging to the first class turns out to be:
\begin{equation}
S_g^{(1)} = (2k + 1)(kf + 1){\Lambda _{g - 1}}.
\end{equation}

The procedure is straightforwardly extended to the other two cases. From the relations of coefficients in Chebyshev polynomials obtained in Eq.~\eqref{Cheb_coeff}, we get the $A^{(2)}(0)$ and $A^{(2)}(1)$ coefficients for the polynomial of Eq.~\eqref{eq_2nd_class} for the second class:
\begin{equation}
\begin{array}{l}
 A^{(2)}(0) = {\alpha _k} - {\alpha _{k - 1}} = 1 \\
 A^{(2)}(1) = {\beta _k} - {\beta _{k{\rm{ - 1}}}} =  - \frac{1}{2}({k^2} + k)
 \end{array}
 \end{equation}
Accounting for the multiplicity $\Delta_g$ [Eq.~\eqref{egDelta}] of Eq.~\eqref{eq_2nd_class} the sum of all inverse eigenvalues in the the second class reads:
\begin{equation}
S_g^{(2)} = \frac{1}{2}({k^2} + k){\Delta _g} = \frac{1}{2}({k^2} + k)[(f - 2){(f + 1)^{g - 1}} + 1].
\end{equation}

Finally, in the third class, the first two coefficients of ${Q_{f,k}}(\lambda )/\lambda$ are:
\begin{equation}
\begin{array}{l}
 A^{(3)}(0) = (f - 1)k + k + 1 = fk + 1 \\
 A^{(3)}(1) = \frac{1}{6}(f - 1)(k - {k^3}) - \frac{1}{6}({k^3} + 3{k^2} + 2k)
 \end{array}
 \end{equation}
and the sum all inverse nonvanishing eigenvalues corresponding to the third class turns to be:
\begin{equation}
S_g^{(3)} =  - \frac{A^{(3)}(1)}{A^{(3)}(0)} = \frac{{f{k^3} + 3{k^2} + (3-f)k}}{{6(fk + 1)}}
 \end{equation}

Summarizing all three classes, the sum of all inverse nonvanishing eigenvalues for EVF is given by ($g\geq2$)
\begin{align}\nonumber
S_g &= S_g^{(1)} + S_g^{(2)} + S_g^{(3)} \\\nonumber
&= \frac{(f + 1)^{g - 2}}{2}\left\{ \frac{f(2k + 1)(kf + 1)(3f-2)}{(3f+2)}\,\,3^{g - 1}\right.\\\nonumber
&\left.-[(f-1){k^2} + k +1](f - 2)\right\}  \\
&+\frac{k(k+1)[(4k-1)f+6]}{6(fk + 1)} - \frac{(2k + 1)(kf + 1)(f+2)}{(f+1)(3f+2)}\label{S_g}
\end{align}

\begin{figure}[t]
\centering
\includegraphics[width=1\linewidth,trim=0 0 0 0]{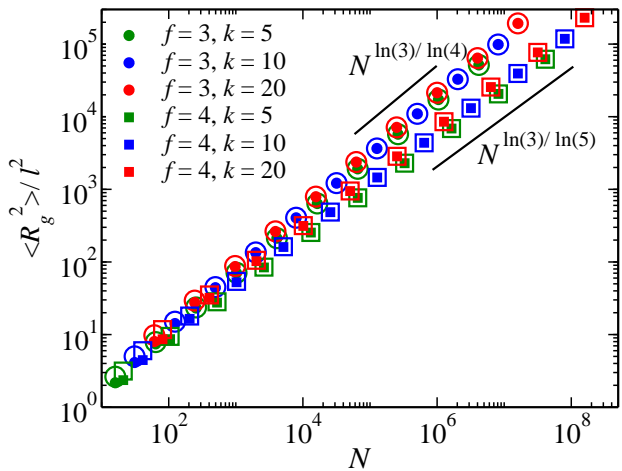}
\caption{Gyration radius of EVF as a function of total number of beads $N$ for different values of the parameters $f$ and $k$. Filled symbols represent the results based on the exact calculations and open symbols show the result of the approximate Eq.~\eqref{RgEVF}. See text for details.}\label{figRg}
\end{figure}

Let us now exemplify the fundamental result of Eq.~\eqref{S_g} by looking at the radius of gyration $\langle R_g^2\rangle$ (Eq.~\eqref{Rg}). To analyze the asymptotic behavior of $\langle R_g^2\rangle$, we look at very high $g$, so that the first term of RHS of Eq.~\eqref{S_g} dominates. With this, the $\langle R_g^2\rangle$ behaves approximately as
\begin{equation}\label{RgEVF}
\frac{\langle R_g^2\rangle}{\ell^2}=\frac{S_g}{N}\simeq\frac{f(3f-2)(2k + 1)}{6(3f+2)(f+1)}\,\,3^g.
\end{equation}
We note that for $f=2$ the RHS of Eq.~\eqref{RgEVF} leads to the well-known value $N/6$ for a linear chain \cite{doi88} and for $k=1$ and $f>2$ it yields the asymptotic $\langle R_g^2\rangle$ of a normal VF. Furthermore, we can rewrite $3^g=(f+1)^{g\ln(3)/\ln(f+1)}$, so that the behavior $\langle R_g^2\rangle\sim N^{\ln(3)/\ln(f+1)}$ readily follows, in accordance with the fractal dimension of the VF in three-dimensional space under the theta-condition, $d_F^{\mathrm{\,3D}}=2\ln(f+1)/\ln 3$ \cite{blumen04}. Thus, for $f=3$ and $f=4$ one has $d_F^{\mathrm{\,3D}}=4\ln(2)/\ln 3\approx2.52372$ and $d_F^{\mathrm{\,3D}}=2\ln(5)/\ln 3\approx2.92995$, respectively, making the structures of $f\leq4$ to be readily embeddable in the three-dimensional space.

In Fig.~\ref{figRg} we compare the exact calculations of $\langle R_g^2\rangle$ for EVF with the asymptotic Eq.~\eqref{RgEVF}. As can be inferred from the figure, Eq.~\eqref{RgEVF} performs well for large $N$. Moreover, one can readily recognize the scaling  $\langle R_g^2\rangle\sim N^{\ln(3)/\ln(f+1)}$.

\section{Dynamics of EVF under external forces}\label{dynamics}

In the GGS formalism \cite{gurtovenko05}, the Laplacian matrix $\mathbf{L}=(L_{ij})$ is directly related to the polymer dynamics. Here the motion of a bead (say, $m$th) located at $\mathbf{R}_m(t)=(X_m(t),Y_{m}(t),Z_{m}(t))$ under an external force $\mathbf{F}_m(t)$ obeys a set of Langevin equations:
\begin{equation}\label{Langevin}
\zeta \frac{d \mathbf{R}_m(t)}{dt} + K\, \sum_{i=1}^N L_{mi}\mathbf{R}_i(t) = \mathbf{w}_m(t) + \mathbf{F}_m(t)\,,
\end{equation}
where $\zeta$ it the friction coefficient (i.e., the related term represents dumping), $K=3k_BT/\ell^2$ is the spring constant, and $\mathbf{w}_m(t)$ is a fluctuating force obeying white noise relations $\langle \mathbf{w}_m(t)\rangle=0$ and $\langle w_{m \alpha}(t)w_{n \beta}(t')\rangle=2k_B T\zeta\delta_{\alpha \beta}\delta_{mn}\delta(t-t')$ ($\alpha$ and $\beta$ denote Cartesian components). We note that formally in Eq.~\eqref{Langevin} the vector $\mathbf{R}_m(t)$ can be of any dimension (only the spring constant $K$ should be then rescaled from $3k_BT/\ell^2$ to $dk_BT/\ell^2$ for dimension $d$). However, in dimension $d=2$ the excluded volume interactions are very strong, so that EVF modelled in the GGS framework get embedding problems even for $f=3$. On the other hand, for higher $d$ the embedding of Gaussian EVF could be readily realized.

Depending on the choice of the external force $\mathbf{F}_m(t)$ in Eq.~\eqref{Langevin}, different physical situations can be considered. We start by looking at the microrheological behavior of EVF, in which the $\mathbf{F}_m(t)$ is applied on a single bead \cite{amblard96}. Let be $\mathbf{F}_m(t) = F\Theta(t)\delta_{mk}\mathbf{e}_y$, i.e., the force acts constantly on $k$th bead in the $y$ direction starting at time $t=0$. The experiment is performed many times by picking randomly EVF's bead. Averaging over all realizations and over the random forces, the bead displacement in the $y$-direction is given by~\cite{schiessel98,biswas00,*kant00,*biswas01,*katyal15}
\begin{equation}\label{Displacement}
\langle Y(t) \rangle = \frac{F t}{N \zeta} + \frac{F\tau_0}{N \zeta} \sum_{i=2}^N \frac{1 -\exp(-\lambda_i^{(g)} t/\tau_0)}{\lambda_i^{(g)}}\,,
\end{equation}
 where $\tau_0=\zeta/K$ is the monomeric relaxation time and $\{\lambda_i^{(g)}\}$ are the nonvanishing eigenvalues  of the Laplacian matrix $\mathbf{L}$ (i.e., all eigenvalues excluding the eigenvalue $\lambda_1^{(g)}=0$).

\begin{figure}[t]
\centering
\includegraphics[width=1\linewidth,trim=0 0 0 0]{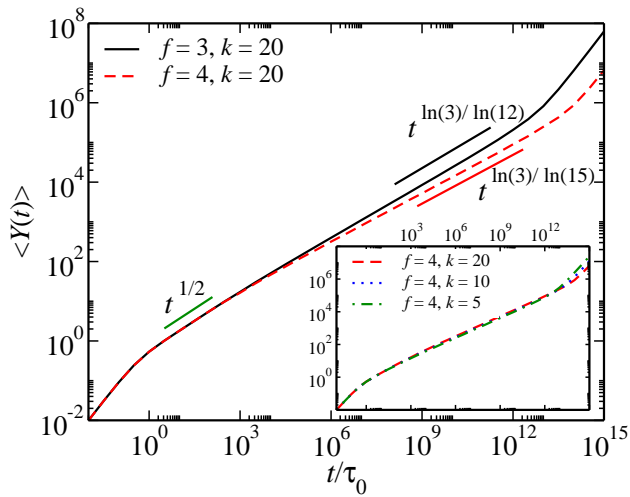}
\caption{Structure-averaged bead displacement $\langle Y(t) \rangle$ under a constantly acting force for EVF of generation $g=10$, spacer parameter $k=20$, and different functionalities $f$. The inset represents the curves for different values of the spacer length $k$. The ratio between parameters $F$ and $\zeta$ is $F/\zeta=1$. See text for details.}\label{Y}
\end{figure}

In Fig.~\ref{Y} we plot the bead displacements $\langle Y(t) \rangle$ for EVF of different functionalities $f$. As can be observed, for short times there is no difference between the $\langle Y(t) \rangle$. In this time domain the intrachain part of spectra becomes apparent (in analogy with  intrachain spectrum in networks \cite{gurtovenko98}), which is partly represented by the second class of eigenvalues, see Eq.~\eqref{lambda2nd}. Hence we see the chain motion, which possesses a typical Rouse behavior $t^{1/2}$ \cite{biswas00}, and do not feel the fractal structure yet. Going to
higher times the fractal structure of EVF becomes evident, we observe a subdiffusive motion $t^{1-\frac{d_s}{2}}$ with the distinct spectral dimension $d_s$ of normal VF, which for different functionalities $f$ is given by $d_s=2\ln(f+1)/\ln(3f+3)$~\cite{blumen04}.

\begin{figure}[t]
\centering
\includegraphics[width=1.05\linewidth,trim=0 0 0 0]{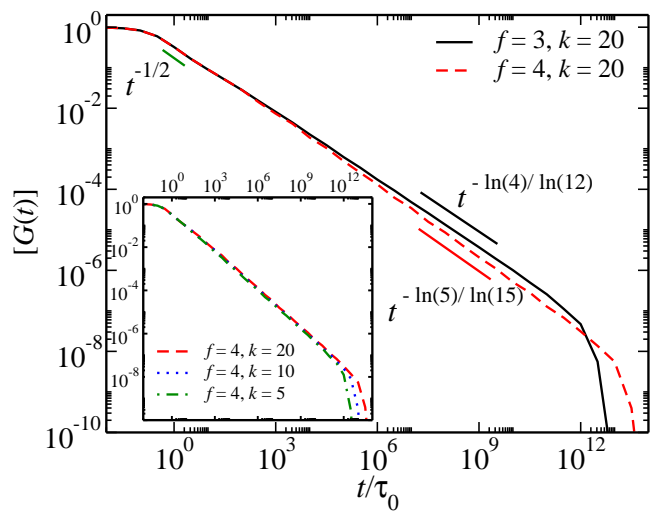}
\caption{Dynamical shear modulus $[G(t)]$ for EVF of generation $g=10$, spacer parameter $k=20$, and different functionalities $f$. The inset represents the curves for different values of the spacer length $k$. See text for details.}\label{Gt}
\end{figure}

Next, we look at the macroscopic rheological properties of EVF. In this case the external force is applied to the whole sample in an oscillating way (with the frequency $\omega$), $\mathbf{F}_m(t) = \gamma_0e^{\mathrm{i}\omega t}Y_m(t)\mathbf{e}_x$. The response to such mechanical impact is represented by the dynamical shear modulus $G(t)$ \cite{doi88,gurtovenko05}, whose normalized form $[G(t)]\equiv G(t)/G(0)$ is given by \cite{gurtovenko02,gurtovenko05}:
\begin{equation}
[G(t)]=\frac{1}{N}\sum_{i=2}^N \exp(-\lambda_i^{(g)} t/\tau_0).
\end{equation}

In Fig.~\ref{Gt} we plot the dynamical modulus $[G(t)]$ for EVF considered in Fig.~\ref{Y}. Also here, for short times one observes intrachain behavior $[G(t)]\sim t^{-1/2}$ \cite{gurtovenko98,gurtovenko02}, which does not reflect the fractal structure of EVF. The situation changes for higher times, where the scaling $[G(t)]\sim t^{-d_s/2}$ clearly distinguishes between EVF of different functionalities $f$.

Finally, we make a remark on the role of spacer length $k$, see the insets to Figs.~\ref{Y} and \ref{Gt}. As can be inferred from the insets, the variation of $k$ (here for $k\geq5$) does not change the characteristic scalings. However, as discussed in Sec.~\ref{spectra} and shown in Table~\ref{table}, increasing the spacer length leads to a decrease of the minimal nonvanishing eigenvalue. Thus, the corresponding maximal relaxation time of the system gets longer and the crossover to the terminal regime takes place at longer times for longer $k$.

\section{Conclusions}\label{conclusions}

Summarizing, in this paper we have studied the Laplacian spectra of the extended Vicsek fractals (EVF). The analytic recursion relations for the spectra allowed us to obtain for EVF an exact analytic expression for the sum of all nonvanishing eigenvalues. This quantity is fundamental for many characteristics: for polymers it describes the gyration radius, the zero shear viscosity, or the Wiener index; for the theory of random walks it is very important for the mean-first passage problems \cite{zhang10,benichou14,chupeau15}.

Introduction of the linear spacers leads to heterogeneities in the fractal behavior, which clearly manifest themselves in the dynamical properties, as we have shown for the structure-averaged bead displacement and for the dynamic shear modulus. With this, our study provides a useful model system for studying heterogeneous fractals or multifractals.

\section*{Acknowledgments}
M.D. thanks Shanghai Key Laboratory of Intelligent Information Processing for hospitality. M.D. acknowledges DFG through GRK 1642/1. Z.Z. was supported by the National Natural Science Foundation of China under Grants No. 11275049.


%

\end{document}